\documentclass[12pt]{article}
\usepackage{graphicx}
\title{A radioactive decay simulation (for Education)}
\author{Ihab F. Riad, Mohammed El Shazali Sir Elkatim,\\
Physics Dept. Faculty of Science, University of Khartoum}
\date{}
\begin{document}
\maketitle
 \begin{abstract} This article is intended for physics
educators and students at school and undergraduate level. It is
used at our department to introduce students to simulation and
offer a guide in using statistics in physics. The simulation code
was created using Matlab, and was given a friendly interface with
a Labview module.
\end{abstract}
\section{ Introduction:}
 During our teaching for undergraduates of physics we saw a need
 for an experiment that can demonstrate the close ties between
 experiments and statistical analysis. We also felt the need of introducing
 the idea of simulating a physics experiment for the huge and increasing role of simulation in the
 different fields of science. We used to have  an
 experiment targeting the Poisson distribution and Radioactivity,
 the experiment was performed with a Geiger counter, so it was a
 good choice for combining the two ideas. The simulation code was
 created with a Matlab script and was then given a friendly
 interface with the help of a Labview module. For running the code
 you need to have Matlab installed on your system.
\newpage
 \section{Theory:}
 The simulation depends on the Normal number generator available
 through the Matlab. To simulate the decay behavior of a
 radioactive sample having $N_o$ nuclide, and decay constant
 $\lambda$ during a time $\Delta t$, $N_o$ uniform random numbers
 were generated, with each number representing a potential decaying
 nuclide. To demonstrate this note that the probability of having
 a random number $x$ between $a$ and $b$ such that $x_o>x>b$ is
 given by
 \begin{eqnarray}
P(x>x_o)&=&\int_{x_o}^b \frac{1}{b-a}dx; \nonumber \\
&=& \frac{(b-x_o)}{b-a};
\end{eqnarray}
setting $a = 0$ and $b = 1$, in our case we get
\begin{equation}
P(x>x_o)=1-x_o.
\end{equation}
Noting that the probability of a single nuclei to decay in a time
$\Delta t$ is $\lambda \Delta t$, or
\begin{eqnarray}\label{condition}
\lambda \Delta t &=& 1- x_o, \quad \textrm{hence} \nonumber \\
x_o &=& 1-\lambda \Delta t.
\end{eqnarray}
i.e any uniform random number grater than $x_o$ represent a
decaying nuclide. The Matlab uniform random number generator is
used to create $N_o$ random number and each is checked with
condition equation(\ref{condition}) to identify it as a
non-decaying or decaying nuclei which in such a case counted.

The time $\Delta t$ is controlled trough the Labview interface. In
the simulation you are given  a number of different samples to
choose among each represent a different decay constant. You also
have three choices of $N_o$ such that for\\
 small $N_o =10^6$, medium $N_o =5 \times 10^6$ and large $N_o =50 \times
 10^6$. The above choices were made for the limited computational
 power of the system used. The number of trials to be simulated
 can be controlled and up to a thousand
 trials can be simulated. The result of each trial is recorded at a .txt file of
 your choice.
\newpage
 \section{Application:}
 The Labview is started, and after making your choices of the
 different variables press the start bottom. The result of each
 trial will be recorded at a file, which can then be read and the
 data treated with a data analysis tool. An example of the
 simulation is the histogram below  created from 851 trials for
 sample {\bf E} with medium number of nuclei for a time of 60min for each trial.
 Using all this information the half life of the sample can be
 calculated
 \begin{equation}
 \bar{N
 }=N_o\left(1-e^{-\lambda \Delta t}\right).
 \end{equation}
 Using the value of $\bar{N}$ collected from the simulation,
 $\lambda$  can be found as
 $$
 79.189 = 5\times 10^6\left(1-e^{-\frac{\lambda}{365 \times
 24}}\right), \nonumber \\
$$
$$
T_{\frac{1}{2}} = 4.990 years.
$$
Compare this to the value set in the code.
\\
\\
\section*{Note:}
If you would like to have the simulation please send me an email
and we will be happy to send it. \clearpage
\begin{figure}[h]
\begin{center}
\includegraphics[scale=1.5]{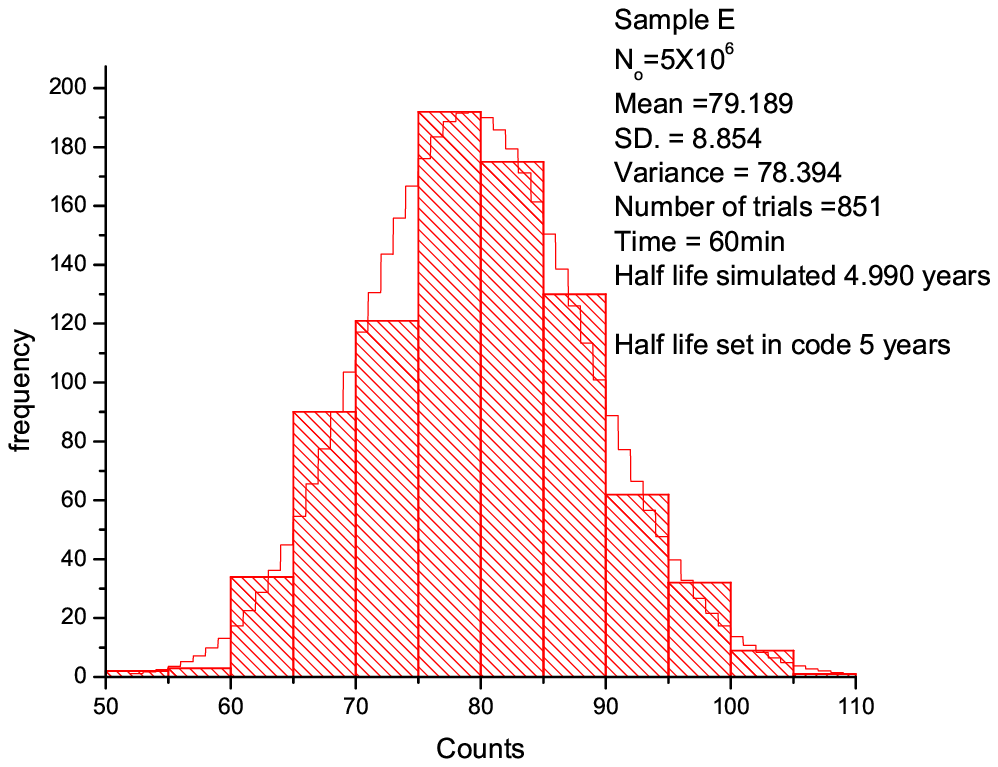}
\end{center}
\end{figure}
\clearpage
 \section{The Simulation:}
 In this section we include the code and a snapshot of the Labview
 interface.\\
 {\bf Matlab Script}\\
 \\
 1: format long g
2: if n==0; \\
3: x= 1-(log(2)/(0.5*24*3600))*1*60*t; \\
4: elseif n==1; \\
5: x= 1-(log(2)/(10*24*3600))*60*t; \\
6: elseif n==2; \\
7: x= 1-(log(2)/(160*24*3600))*60*t;\\
 8: elseif n==3; \\
 9: x=1-(log(2)/(2*365*24*3600))*60*t; \\
 10: elseif n==4; \\
 11: x=1-(log(2)/(5*365*24*3600))*60*t; \\
 12: end \\
 13: if c==0; \\
 14:d=1000000; \\
 15: elseif c==1;\\
  16: d=5000000; \\
  17: elseif c==2;\\
   18:d=50000000; \\
   19: end \\
   20: z=0; \\
   21: for i=1 : d \\
   22: y=rand; \\
   23: if y>x \\
   24: z=z+1; \\
   25: else \\
   26: z=z; \\
   27: end \\
   28:end \\
   29: z

 \clearpage
\begin{figure}[h]
\begin{center}
\includegraphics[scale=1.5]{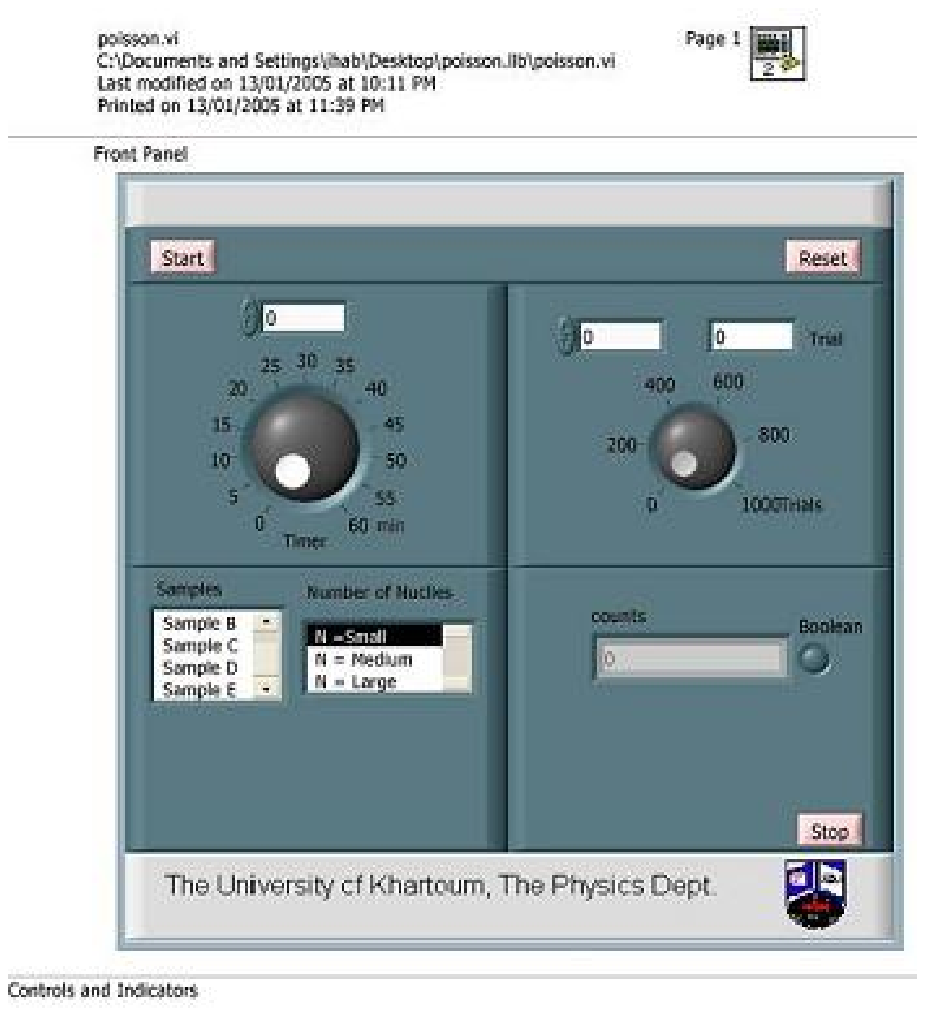}
\end{center}
\end{figure}

\end{document}